\documentclass[letterpaper,12pt,notoc]{JHEP3}

\def\mc{\mathcal}

\usepackage{graphicx}
\usepackage{amsmath}
\usepackage{amssymb}

\preprint{ \hbox{}\hfill arXiv: 1206.2150}

\title{Gaugings of $N=4$ three dimensional gauged
supergravity with exceptional coset manifolds}
\author{Parinya Karndumri\\
String Theory and Supergravity Group, Department
of Physics, Faculty of Science, Chulalongkorn University, 254 Phayathai Road, Pathumwan, Bangkok 10330, Thailand\\
Thailand Center of Excellence in Physics, CHE, Ministry of Education, Bangkok 10400, Thailand \\
E-mail: \email{parinya.ka@hotmail.com}}

\abstract{Some admissible gauge groups of $N=4$ Chern-Simons gauged
supergravity in three dimensions with exceptional scalar manifolds
$G_{2(2)}/SO(4)$, $F_{4(4)}/USp(6)\times SU(2)$,
$E_{6(2)}/SU(6)\times SU(2)$, $E_{7(-5)}/SO(12)\times SU(2)$ and
$E_{8(-24)}/E_7\times SU(2)$ are identified. In particular, a
complete list of all possible gauge groups is given for the theory
with $G_{2(2)}/SO(4)$ coset space. We also study scalar potentials
for all of these gauge groups and find some critical points. In the
case of $F_{4(4)}/USp(6)\times SU(2)$ target space, we give some
semisimple gauge groups which are maximal subgroups of $F_{4(4)}$.
Most importantly, we construct the $SO(4)\ltimes \mathbf{T}^6$
gauged supergravity which is equivalent to $N=4$ $SO(4)$ Yang-Mills
gauged supergravity. The latter is proposed to be obtained from an
$S^3$ reduction of $(1,0)$ six dimensional supergravity coupled to
two vector and two tensor multiplets. The scalar potential of this
theory on the scalar fields which are invariant under $SO(4)$ is
explicitly computed. Depending on the value of the coupling
constants, the theory admits both dS and AdS vacua when all of the
28 scalars vanish. The maximal $N=4$ supersymmetric $AdS_3$ should
correspond to the $AdS_3\times S^3$ solution of the $(1,0)$ six
dimensional theory. Finally, some gauge groups of the theories with
$E_{6(2)}/SU(6)\times SU(2)$, $E_{7(-5)}/SO(12)\times SU(2)$ and
$E_{8(-24)}/E_7\times SU(2)$ scalar manifolds are identified.}

\begin{document}
\section{Introduction}
Three dimensional Chern-Simons (CS) gauged supergravity has a very
rich structure \cite{nicolai1}, \cite{N8}. Many possible gauge
groups of various types are allowed \cite{nicolai2},
\cite{nicolai3}. This is due to the duality between scalars and
vectors in three dimensions. The propagating bosonic degrees of
freedom in the ungauged theory are given by scalar fields. The gauge
fields enter the gauged Lagrangian via Chern-Simons terms which do
not introduce any extra degrees of freedom. So, unlike in higher
dimensional analogues, the dimensions of possible gauge groups are
not restricted by the number of vector fields present in the
ungauged theory.
\\
\indent Of particular interest are non-semisimple gauge groups of
the form $G_0\ltimes \mathbf{T}^{\textrm{dim}\, G_0}$. This is
on-shell equivalent to a Yang-Mills (YM) gauged supergravity with
gauge group $G_0$ usually obtained from a dimensional reduction of
some higher dimensional supergravity \cite{csym}. Working with the
CS gauged supergravity is much simpler than the equivalent YM type
theory. This has been emphasized in \cite{csym} in which the
comparison between a simple Lagrangian of the CS type supergravity
and a much more complicated Lagrangian of the YM type supergravity
has been pointed out. Therefore, the CS construction is more
convenient to work with in a three dimensional framework.
\\
\indent The CS gauged supergravity is generally described by
gaugings of the ungauged theory in the form of a non-linear sigma
model coupled to supergravity \cite{dewit1}. While pure supergravity
in three dimensions is topological, for some earlier construction of
CS three dimensional supergravity see \cite{Deser}, this is not the
case for the matter coupled supergravity. The target space for
scalar fields in the theory with $N>4$ is a symmetric space of the
form $G/H$ in which the global and local symmetry groups are given
by $G$ and $H$, respectively. All symmetric spaces involved in
$N=5,6,8,9,10,12,16$ have been given in \cite{dewit1}. In this
paper, we will focus on the $N=4$ theory whose scalar manifold is
generally a product of two (not necessarily symmetric) quaternionic
manifolds. Furthermore, we are interested in the case of symmetric
target spaces and, in particular, the so-called degenerate case in
which the target space contains only one quaternionic manifold.
\\
\indent With the embedding tensor formalism of \cite{nicolai1},
gaugings can be studied in a $G$ covariant manner. This is very
useful, particularly, in the case of symmetric target spaces in
which the classification of gauge groups can be achieved by group
theoretical method. In this formalism, the symmetric and gauge
invariant embedding temsor $\Theta$ is introduced. It acts as a
projector on the global symmetry group $G$ to the corresponding
gauge group $G_0\subset G$. In defining a consistent gauge group,
the associated embedding tensor needs to satisfy quadratic and
linear constraints coming from the closure of the gauge algebra and
the consistency with supersymmetry. The general formulation of this
gauged supergravity for any value of $N$ has been constructed in
\cite{dewit}. Some semisimple gauge groups for theories with $N>4$
have also been given.
\\
\indent Gaugings of CS three dimensional gauged supergravity with
different value of $N$ have been studied in various places
\cite{nicolai1}, \cite{nicolai2}, \cite{nicolai3}, \cite{N6},
\cite{henning_AdS3_S3}, \cite{henning_hohm} and \cite{bs}. In
\cite{gkn}, \cite{KK} and \cite{rg_int}, non-semisimple gaugings of
the $N=4$ theory with scalar manifold $SO(4,k)/SO(4)\times SO(k)$
have been studied. The higher dimensional origin for some class of
gaugings has also been identified. In the following, we will study
$N=4$ theory with exceptional coset manifolds and identify some of
their gauge groups for both semisimple and non-semisimple types. All
the exceptional coset manifolds for the $N=4$ theory have been given
in \cite{dewit}. These are $G_{2(2)}/SO(4)$, $F_{4(4)}/USp(6)\times
SU(2)$, $E_{6(2)}/SU(6)\times SU(2)$, $E_{7(-5)}/SO(12)\times SU(2)$
and $E_{8(-24)}/E_7\times SU(2)$.
\\
\indent Since $G_{2(2)}$ is a small group, we can give all of the
possible gauge groups in a theory with $G_{2(2)}/SO(4)$ coset
manifold. We will also study scalar potentials of these gauge
groups. The ungauged version of this theory can be obtained by a
$T^2$ reduction of the minimal five dimensional supergravity, see
for example \cite{G2_from5D1} and \cite{G2_from5D2}. The $S^2$
reduction of this five dimensional theory should give the gauged
version in three dimensions with $SO(3)\ltimes \mathbf{T}^3$ gauge
group as proposed in \cite{csym}. But, as we will see, it can be
verified that this gauge group cannot be embedded in $G_{2(2)}$. So,
if the corresponding $S^2$ reduction exists, it is very interesting
to find a description in term of the three dimensional CS framework.
\\
\indent In the $F_{4(4)}/USp(6)\times SU(2)$ case, we will give some
admissible semisimple gauge groups and a non-semisimple gauge group
$SO(4)\ltimes \mathbf{T}^6$. The latter is one of the interesting
results in this paper and should describe an $S^3$ reduction of
$(1,0)$ six dimensional supergravity coupled to two vector and two
tensor multiplets. Dimensional reductions of this six dimensional
theory have been studied before. Firstly, the $SU(2)$ reduction of
pure supergravity in six dimensions has been investigated in
\cite{PopeSU2} and \cite{PopeSU22}. This gives rise to $SO(3)$ YM
gauged supergravity in three dimensions coupled to three massive
vector fields. The theory is in turn equivalent to $SO(3)\times
(\mathbf{T}^3,\hat{\mathbf{T}}^3)$ CS gauged supergravity as
proposed by \cite{csym} with the three nilpotent generators of
$\hat{\mathbf{T}}^3$ describing the three massive vector fields.
Then, the $SU(2)$ reduction of the same theory coupled to an
anti-symmetric tensor multiplet and $\textrm{dim}G_c$ vector
multiplets of any semisimple gauge group $G_c$ has been constructed
in \cite{KK}. The resulting three dimensional theory is $SO(3)\times
G_c$ YM gauged supergravity without any massive vector fields since
they are truncated out in the process of reduction.
\\
\indent In this paper, we give one more example of the possible
reduction of this six dimensional supergravity coupled to two vector
and two tensor multiplets on an $S^3$. The reduction on $T^3$ giving
rise to the ungauged three dimensional supergravity with
$F_{4(4)}/USp(6)\times SU(2)$ has been studied in details in
\cite{Pope_3D_coset}. The reduction on $S^3$ would give the gauged
version of the three dimensional theory with $SO(4)$ or $SO(4)\times
\mathbf{T}^6$ for YM and CS supergravities, respectively. We propose
the corresponding reduction by constructing the equivalent CS gauged
supergravity directly in three dimensions. As mentioned above, the
CS version is much simpler to deal with, so this is a convenient
starting point. Moreover, both the six dimensional theory and the
resulting three dimensional one are in the so-called magical
supergravities whose gaugings have been studied recently in
\cite{henning_magic}.
\\
\indent We end the paper by giving some admissible gauge groups in
the remaining cases with $E_{6(2)}/SU(6)\times SU(2)$,
$E_{7(-5)}/SO(12)\times SU(2)$ and $E_{8(-24)}/E_7\times SU(2)$
scalar manifolds. Apart from the $G_{2(2)}/SO(4)$ case, the lists of
gauge groups identified in this paper are by no means complete.
Further studies are needed in order to cover a larger class of
possible gauge groups.
\\
\indent The paper is organized as follow. In section \ref{N4theory},
we review the $N=4$ three dimensional gauged supergravity with
symmetric target spaces. The consistency conditions on the embedding
tensor are given. We then focus on a specific case of exceptional
cosets. Gaugings for each exceptional symmetric space are identified
in section \ref{G2}, \ref{F4} and \ref{E678}. We also study scalar
potentials for all gauge groups in the $G_{2(2)}/SO(4)$ case and
$SO(4)\ltimes \mathbf{T}^6$ gauging in the $F_{4(4)}/USp(6)\times
SU(2)$ case. The computations are carried out by using the computer
program \textsl{Mathematica}. We finally give some conclusions and
comments in section \ref{conclusions}. Additionally, there is one
appendix in which some useful formulae can be found.
\section{$N=4$ gauged supergravity in three
dimensions}\label{N4theory} In this section, we review the
construction of three dimensional gauged supergravities following
\cite{dewit}. We will repeat only relevant information to describe
$N=4$ theory and refer the reader to \cite{dewit} for the full
details.
\\
\indent Three dimensional ungauged supergravities are described by
nonlinear sigma models coupled to supergravity. Coupling to $N$
extended supergravity requires the existence of $N-1$ hermitean
almost complex structures, $f^P$, $P=2,\ldots N$, on the target
space of the sigma model. The $f^{IJ}$, $I,J=1,\ldots N$, tensors
constructed from $f^P$ via
\begin{equation}
f^{1P}=-f^{P1}=f^P,\qquad f^{PQ}=f^{[P}f^{Q]}\label{fIJ}
\end{equation}
are generators of $SO(N)$ R-symmetry. For $N=4$ theory, the tensor
$J=\frac{1}{6}\epsilon_{PQR}f^Pf^Qf^R$ commutes with the almost
complex structures and satisfies
\begin{equation}
J^2=\mathbf{1},\qquad J_{ij}=J_{ji},\qquad
J=\frac{1}{24}\epsilon^{IJKL}f^{IJ}f^{KL}\, .
\end{equation}
Here and from now on, indices $i,j=1\ldots d$ label coordinates on
the target space whose dimension is $d$. We will also use the flat
target space indices $A,B,\ldots$ with the vielbein defined in the
appendix. The product structure of the target space is implied by
the fact that $J$ is covariantly constant. So, in general, the
target spaces of $N=4$ three dimensional gauged supergravity are
products of two quaternionic spaces. The dimension of the target
space is thus given by $d=d_++d_-$ with $d_\pm$ being the dimensions
of the corresponding two subspaces. Each subspace describes one of
the inequivalent multiplets.  Although, in three dimensions, there
are inequivalent supermultiplets for $N=4$ mod 4, the requirement
that the maximal compact subgroup $SO(N)\times H'\subset G$ must act
irreducibly on the target space allows only one type of the
multiplets \cite{dewit1}. This is not the case for $N=4$ because the
R-symmetry $SO(4)$ is not a simple group and decomposes according to
$SO(4)\sim SO(3)\times SO(3)$, and each factor acts on the two
subspaces, separately. So, $N=4$ theory is special in many aspects
compared to theories with other values of $N$.
\\
\indent Unlike theories with $N>4$, the scalar manifolds of the
$N=4$ theory need not be symmetric. However, in this work we are
interested in the case of symmetric target spaces of the form $G/H$
in which the maximal compact subgroup $H$ contains the R-symmetry
$SO(N)$ or $H=SO(N)\times H'$. As explained above, for the $N=4$
theory, we have $H=SO(3)\times H'\sim SU(2)\times H'$ for each
subspace of the full target space. In this work, we are particularly
interested in the exceptional coset spaces which are of the form
``non-compact real form of some exceptional group/maximal compact
subgroup''. They are quaternionic manifolds $G_{2(2)}/SO(4)$,
$F_{4(4)}/USp(6)\times SU(2)$, $E_{6(2)}/SU(6)\times SU(2)$,
$E_{7(-5)}/SO(12)\times SU(2)$ and $E_{8(-24)}/E_7\times SU(2)$.
Furthermore, we will consider only the degenerate case namely
$d_+d_-=0$. In this case, there is only one quaternionic manifold in
the target space. The relevant formulae and useful relations for a
symmetric target space are reviewed in the appendix.
\\
\indent Gaugings are implemented by promoting some isometries of the
target space to a local symmetry. This procedure is easily dealt
with by introducing the so-called embedding tensor,
$\Theta_{\mathcal{MN}}$, which is gauge invariant. In order to
describe a consistent gauging, the embedding tensor needs to satisfy
two consistency conditions. The first condition called the quadratic
constraint is imposed by the requirement that the gauge generators
$J_{\mathcal{M}}=\Theta_{\mathcal{MN}}t^{\mathcal{N}}$ with
$t^{\mc{N}}$ being generators of $G$ form an algebra. This
constraint is given by \cite{dewit}
\begin{equation}
\Theta_{\mathcal{PL}}f^{\mathcal{KL}}_{\phantom{asds}\mathcal{(M}}\Theta_{\mathcal{N)K}}=0\,
.\label{theta_quadratic}
\end{equation}
Furthermore, in order to be consistent with supersymmetry, there is
a projection constraint on the T-tensor
\begin{equation}
T^{IJ,KL}=T^{[IJ,KL]}-\frac{4}{N-2}\delta^{[I[K}T^{L]M,MJ]}-\frac{2}{(N-1)(N-2)}\delta^{I[K}\delta^{L]J}T^{MN,MN},\label{Tconstraint}
\end{equation}
or equivalently,
\begin{equation}
\mathbb{P}_\boxplus T^{IJ,KL}=0\, .\label{Tconstraint1}
\end{equation}
The T-tensor itself is defined by the image of the embedding tensor
under the map $\mc{V}$
\begin{equation}
T_{\mathcal{AB}}=\mathcal{V}^{\mc{M}}_{\phantom{as}\mc{A}}\Theta_{\mc{MN}}\mathcal{V}^{\mc{N}}_{\phantom{as}\mc{B}}\label{T_tensor_def}
\end{equation}
with the index $\mathcal{A}=\{IJ,\alpha,A\}$. The linear constraint
means that supersymmetry requires the absence of the $\boxplus$
representation of $SO(N)$ in the T-tensor. The projector
$\mathbb{P}_\boxplus$ projects the $SO(N)$ representation on to the
$\boxplus$ representation.
\\
\indent For symmetric target spaces, the condition
\eqref{Tconstraint1} can be lifted to the consistency condition on
the embedding tensor and the map $\mc{V}$, which is now an
isomorphism, can be obtained from the coset representative $L$, see
the relevant formulae in the appendix. The embedding tensor lives in
the symmetric product of the adjoint representation of $G$. This
product can be decomposed into irredeucible representations of $G$
as
\begin{eqnarray}
(R_{\textrm{adj}}\times
R_{\textrm{adj}})_{\textrm{sym}}&=&\mathbf{1}\oplus
\left[\bigoplus_i R_i\right], \nonumber \\
\textrm{or}\qquad\Theta_{\mc{MN}} &=&\theta
\eta_{\mathcal{M}\mathcal{N}}+\sum_i
\mathbb{P}_{R_i}\Theta_{\mathcal{M}\mathcal{N}}\,
.\label{theta_decom}
\end{eqnarray}
$\eta_{\mathcal{M}\mathcal{N}}$ is the Cartan-Killing form of $G$,
and $\mathbb{P}_{R_i}$ is the G-invariant projector onto the
representation $R_i$. Only one representation of $R_i$'s denoted by
$R_0$ contains the $\boxplus$ representation of $SO(N)$ in the
branching under $SO(N)$. The condition \eqref{Tconstraint1} can then
be written as
\begin{equation}
\mathbb{P}_{R_0}T_{\mc{AB}}=0\, .
\end{equation}
Using the fact that this condition is G-covariant, we can write it
as the condition on $\Theta$ \cite{dewit}
\begin{equation}
\mathbb{P}_{R_0}\Theta_{\mc{MN}}=0\, .\label{Theta_constraint}
\end{equation}
\indent For symmetric spaces in the form of exceptional coset
spaces, the embedding tensor takes the simple form
\begin{equation}
\Theta_{\mathcal{M}\mathcal{N}}=\theta
\eta_{\mathcal{M}\mathcal{N}}+\mathbb{P}_{R_1}\Theta_{\mathcal{M}\mathcal{N}}\,
.
\end{equation}
This is because there are only three representations appearing in
the decomposition \eqref{theta_decom} with one of them being the
forbidden representation $R_0$. With this simple structure of the
embedding tensor, we can use group theoretical argument given in
\cite{dewit} to find admissible gauge groups $G_0\subset G$. The
arguments simply says that a semisimple subgroup $G_0\subset G$,
which is a simple group, is admissible if the decomposition of $R_0$
under $G_0$ does not contain a singlet, and a semisimple subgroup
$G_0\subset G$ of the form $G_1\times G_2$ is admissible if the
decomposition of $R_0$ under $G_1\times G_2$ contains precisely one
singlet with a fixed ratio of the coupling constants. In the next
sections, we will use these useful conditions to determine some
admissible gauge groups of the $N=4$ gauged supergravity with
exceptional cosets mentioned above. The relevant group theory
decompositions can be found in \cite{Slansky} and \cite{McKay}.
\\
\indent For conveniences, we also repeat the exceptional
quaternionic spaces together with the decomposition of the
corresponding embedding tensor as well as the representation $R_0$
from \cite{dewit}. \vspace{0.5cm}
\begin{center}
\begin{tabular}{|c|c|c|c|}
  \hline
  $G/H$ & $d$ & $R_{\textrm{adj}}$ & $(R_{\textrm{adj}}\times  R_{\textrm{adj}})_{\textrm{sym}}$\\ \hline
  $\frac{G_{2(2)}}{SO(4)}$ & $8$ & $\mathbf{14}$ & $\mathbf{1}+\mathbf{27}+\underline{\mathbf{77}}$\\ \hline
  $\frac{F_{4(4)}}{USp(6)\times SU(2)}$ & $28$ & $\mathbf{52}$ & $\mathbf{1}+\mathbf{324}+\underline{\mathbf{1053}}$ \\ \hline
  $\frac{E_{6(2)}}{SU(6)\times SU(2)}$ & $40$ & $\mathbf{78}$ & $\mathbf{1}+\mathbf{650}+\underline{\mathbf{2430}}$ \\ \hline
  $\frac{E_{7(-5)}}{SO(12)\times SU(2)}$ & $64$ & $\mathbf{133}$ & $\mathbf{1}+\mathbf{1539}+\underline{\mathbf{7371}}$ \\ \hline
  $\frac{E_{8(-24)}}{E_7\times SU(2)}$ & $112$ & $\mathbf{248}$ & $\mathbf{1}+\mathbf{3875}+\underline{\mathbf{27000}}$ \\
  \hline
\end{tabular}
\end{center}
\textbf{Table1}: Symmetric spaces for $N=4$ supergravity and the
corresponding $R_0$ representation denoted by the underlined
representation.
\section{Gaugings in $G_{2(2)}/SO(4)$ coset manifold}\label{G2}
In this case, the group $G$ is given by $G_{2(2)}$, the split form
of the exceptional group $G_2$. The representation $R_0$ is the
$\mathbf{77}$ representation. The embedding tensor lives in the
representation $\mathbf{1}+\mathbf{27}$. We find that the
appropriate real forms, which can be embedded in $G_{2(2)}$, of
$A_2$ and $A_1\times A_1$ subgroups can be gauged since the
$\mathbf{77}$ representation of $G_2$ contains none and one singlet
when branched under $A_2$ and $A_1\times A_1$, respectively.
\\
\indent From this, we obtain the admissible gauge groups:
\begin{itemize}
  \item $SU(2,1)$
  \item $SL(3,\mathbb{R})$
  \item $SO(4)\sim SU(2)\times SU(2)$
  \item $SO(2,2)\sim SL(2,\mathbb{R})\times SL(2,\mathbb{R})\sim
SO(2,1)\times SO(2,1)$\, .
\end{itemize}
Since the $G_2$ is more tractable, we can also check all other
subgroups whether they can be gauged. First of all, one of the two
$SU(2)$'s in the $SO(4)$ cannot be gauged as well as their diagonal
subgroup $SU(2)_{\textrm{diag}}\subset (SU(2)\times
SU(2))_{\textrm{diag}}$. The $U(1)$ and $U(1)\times U(1)\subset
SU(2)\times SU(2)$ cannot be gauged. The non-semisimple group
$SO(2)\ltimes \mathbf{T}^2$ and the nilpotent $\mathbf{T}^3$, see
below, cannot be gauged either. Therefore, there are no other gauge
groups apart from those listed above.
\\
\indent The split form $G_{2(2)}$ has three commuting generators
\cite{max_ideal}. It has been proposed in \cite{csym} that this
theory with non-semisimple gauge group $SO(3)\ltimes \mathbf{T}^3$
could describe the $S^2$ reduction of the minimal five dimensional
supergravity. It is well-known that the ungauged $N=4$ theory with
coset space $G_{2(2)}/SO(4)$ can be obtained from $N=2$ supergravity
in five dimensions reduced on $T^2$. Unfortunately, the group
$SO(3)\ltimes \mathbf{T}^3$ cannot be embedded in $G_{2(2)}$ as can
be explicitly checked from the generators of $G_{2(2)}$ given in the
next subsection. So, $SO(3)\ltimes \mathbf{T}^3$ is certainly not
admissible. Furthermore, the nilpotent gauging, whose example in
$N=16$ theory has been given in \cite{nicolai3}, with the
corresponding gauge group $\mathbf{T}^3$ cannot be gauged. Finally,
the group $G_{2(2)}$ itself can be gauged but the corresponding
scalar potential will be a cosmological constant. It is the general
fact that the group $G$ is always admissible with a constant scalar
potential. Since the group $G_{2(2)}$ is quite small, the
computation of the corresponding scalar potential for each gauge
group is not so difficult to perform. We then study scalar
potentials and the associated critical points for all of the
admissible gauge groups in the remaining subsections.

\subsection{$SO(4)$ gauging}
We first give the structure of the $G_{2(2)}$ coset. The maximal
compact subgroup is $SO(4)\sim SU(2)\times SU(2)$. Therefore, there
are eight scalars. Under $SO(4)\sim SU(2)\times SU(2)$, they
transform as $(\mathbf{2},\mathbf{4})$. This means that they are
four copies of the spinor representation of the first $SU(2)$. We
then identify this group as the R-symmetry. The other $SU(2)$ would
become the group $H'$ mentioned in section \ref{N4theory}
\\
\indent We will use the explicit generators of $G_2$ given in
\cite{G2 Euler}. The corresponding generators of the split from
$G_{2(2)}$ are given in \cite{exceptional coset}. These generators
are denoted by $Q_i$, $i=1,\ldots, 14$ in \cite{exceptional coset}.
In order to apply our general formalism, we relabel the generators
as follow.
\begin{itemize}
  \item R-Symmetry generators: $T_{12}=\frac{1}{2}Q_3$,
  $T_{13}=-\frac{1}{2}Q_2$, $T_{23}=\frac{1}{2}Q_1$
  \item Non-compact generators:
  \begin{equation}
  Y_A=\left\{
               \begin{array}{ll}
                 \frac{1}{2}Q_{A+3}, & \hbox{A=1,2,3,4}, \\
                 \frac{1}{2}Q_{A+6}, & \hbox{A=5,6,7,8}
               \end{array}
             \right.\, .
  \end{equation}
\end{itemize}
The coset representative can be parametrized by using the Euler
angle parametrization given in \cite{G2 Euler}
\begin{equation}
L=e^{a_1Q_1}e^{a_2Q_2}e^{a_3Q_3}e^{a_4Q_8}e^{a_5Q_9}e^{a_6Q_{10}}e^{b_1Y_2}e^{b_2\sqrt{3}Y_5}\,
.
\end{equation}
We find the embedding tensor to be
\begin{equation}
\Theta=g(\Theta_{SU(2)^{(2)}}-3\Theta_{SU(2)^{(1)}}).
\end{equation}
In terms of the $SO(4)$ gauge generators $Q_i$, $i=1,2,3,8,9,10$,
the generators for the gauge groups, $SU(2)^{(1)}$ and
$SU(2)^{(2)}$, are given by
\begin{itemize}
  \item $SU(2)^{(1)}$: $J^{(1)}_i=\frac{1}{2}Q_i$, $i=1,2,3$,
  \item $SU(2)^{(2)}$: $J^{(2)}_i=\frac{1}{2}Q_{i+7}$, $i=1,2,3$\, .
\end{itemize}
Using the formulae in the appendix, the scalar potential is found to
be
\begin{eqnarray}
V&=&-\frac{27}{2}g^2\left[21+\cosh (4b_1)+8(\cosh b_1+\cosh
(3b_1))\cosh b_2\right . \nonumber \\
& &\left . 4\cosh(2b_2)+2\cosh (2b_1)(10+\cosh (2b_2))\right]
\end{eqnarray}
which does not depend on $a_i$. There is only a trivial critical
point at $b_1=b_2=0$. The critical point preserves $(4,0)$
supersymmetry with the value of the potential at the critical point
$V_0=-1296g^2$. The full isometry of the corresponding $AdS_3$
background is $SU(1,1|2)\times SU(1,1)$. The bosonic subgroup of
$SU(1,1|2)$ is given by $SU(1,1)\times SU(2)$ \cite{SC}. This
symmetry is the same as the $N=4$ superconformal symmetry in the
dual two dimensional CFT in the context of the AdS/CFT
correspondence \cite{maldacena}. The $SU(1,1)$ is a part of the
$AdS_3$ isometry $SO(2,2)\sim SO(2,1)\times SO(2,1)\sim
SU(1,1)\times SU(1,1)$ while the $SU(2)$ is the R-symmetry. The
eight supercharges transform as
$(\mathbf{2},\mathbf{2})+(\mathbf{2},\mathbf{2})$ under
$SU(1,1)\times SU(2)$.

\subsection{$SO(2,2)$ gauging}
For the gauge group $SO(2,2)\sim SL(2,\mathbb{R})\times
SL(2,\mathbb{R})$, the generators are given by
\begin{itemize}
  \item $SL(2,\mathbb{R})^{(1)}$: $j^{(1)}_1=\frac{1}{2}Q_4$,
  $j^{(1)}_2=\frac{1}{2}Q_5$, $j^{(1)}_3=\frac{1}{4}(Q_3+\sqrt{3}Q_8)$
  \item $SL(2,\mathbb{R})^{(2)}$: $j^{(2)}_1=\frac{1}{2}Q_{11}$,
  $j^{(2)}_2=\frac{1}{2}Q_{12}$, $j^{(2)}_3=\frac{1}{4}(Q_3-\frac{1}{\sqrt{3}}Q_8)$\, .
\end{itemize}
Four of the eight scalars are parts of the gauge group while the
remaining four correspond to non-compact generators of another
$SL(2,\mathbb{R})\times SL(2,\mathbb{R})$. The coset representative
can be parametrized by
\begin{equation}
L=e^{a_1\frac{1}{4}(Q_3-\sqrt{3}Q_8)}e^{b_1Y_3}e^{a_2\frac{1}{4}(\sqrt{3}Q_3+Q_8)}e^{b_2Y_7}\,
.
\end{equation}
The embedding tensor is given by
\begin{equation}
\Theta=g(\Theta_{SL(2)^{(2)}}-3\Theta_{SL(2)^{(1)}}).
\end{equation}
Notice that the ratio of the coupling constants of the two factors
is the same as in the $SO(4)$ case. This emphasizes the fact that
the ratio of the two couplings does not depend on the different real
forms of the gauge group \cite{nicolai2}.
\\
\indent The resulting scalar potential is given by
\begin{eqnarray}
V&=&-\frac{3}{2}g^2\left[21+4\cosh(2b_1)-40\cosh b_1\cosh\frac{b_2}{\sqrt{3}}  \right.\nonumber \\
& &\left.
+2(10+\cosh(2b_1))\cosh\frac{2b_2}{\sqrt{3}}+\cosh\frac{4b_2}{\sqrt{3}}-8\cosh
b_1\cosh(\sqrt{3}b_2)\right].
\end{eqnarray}
The trivial critical point at $b_1=b_2=0$ is a Minkowski vacuum with
$V_0=0$ and preserves the full $N=4$ supersymmetry. The gauge
symmetry preserved by this critical point is the maximal compact
subgroup of the gauge group $U(1)\times U(1)$.
\\
\indent With the relation $\cosh b_1=2\cosh\frac{b_2}{\sqrt{3}}$,
there are dS vacua with
$V_0=18g^2\cosh^2\frac{b_2}{\sqrt{3}}\cosh\frac{2b_2}{\sqrt{3}}$
depending on the value of $b_2$. For both $b_1$ and $b_2$ non zero,
the critical point beaks all the gauge symmetry. For $b_1=0$ or
$b_2=0$, the $U(1)_{\textrm{diag}}\subset U(1)\times U(1)$ is
preserved. But, $b_1$ cannot be zero since this will give imaginary
$b_2$. So, the $U(1)_{\textrm{diag}}$ point is given by
$b_1=\cosh^{-1}2, b_2=0$ and $V_0=18g^2$.
\subsection{$SU(2,1)$ gauging}
By the construction given in \cite{G2 Euler}, the first eight
matrices $c_i$, $i=1,\ldots, 8$ generate the $SU(3)\subset G_2$. In
the split form $G_{2(2)}$, this corresponds to the $SU(2,1)$
subgroup of $G_{2(2)}$. The gauge generators are then given by
$Q_i$, $i=1,\ldots, 8$ in which the maximal compact subgroup
$SU(2)\times U(1)$ is generated by $\{Q_1,Q_2,Q_3\}$ and $Q_8$,
respectively.
\\
\indent The embedding tensor is given by
\begin{equation}
\Theta=g\eta_{SU(2,1)}
\end{equation}
where $\eta_{SU(2,1)}$ is the Cartan-Killing form of $SU(2,1)$. It
does not seem to be possible to find a simple parametrization of the
four relevant scalars corresponding to $Y_i$, $i=5,6,7,8$.
Therefore, we simply parametrize the coset representative by
\begin{equation}
L=e^{b_1Y_5}e^{b_2Y_6}e^{b_3Y_7}e^{b_4Y_8}\, .
\end{equation}
The potential turns out to be very long and complicated, so we will
not attempt to give its explicit form, here. The trivial critical
point at $b_1=b_2=b_3=b_4=0$ has $(4,0)$ supersymmetry and
$V_0=-16g^2$. The residual gauge symmetry is given by $SU(2)\times
U(1)$. Apart from this point, it is most likely that there are no
other critical points. However, more detailed investigations are
needed in order to have a definite conclusion.
\subsection{$SL(3,\mathbb{R})$ gauging}
The non-compact form $SL(3,\mathbb{R})$ has $SO(3)$ as its maximal
compact subgroup. This subgroup is formed by the diagonal subgroup
of the two $SU(2)$'s in $SO(4)$. Recall that the eight scalars
transform as $(\mathbf{2},\mathbf{4})$ under the $SO(4)$, we find
that under the $SO(3)_{\textrm{diag}}$, the scalars transform as
\begin{displaymath}
\mathbf{2}\times\mathbf{4}=\mathbf{3}+\mathbf{5}\, .
\end{displaymath}
The $\mathbf{5}$ representation forms five non-compact generators
and extends the $SO(3)_{\textrm{diag}}$ to the full
$SL(3,\mathbb{R})$ gauge group. The $\mathbf{3}$ representation
gives the remaining three scalars to be parametrized in the coset
representative $L$.
\\
\indent The $SL(3,\mathbb{R})$ generators are then given by
\begin{eqnarray}
R_1&=&\frac{1}{2}(Q_1+\sqrt{3}Q_9),\qquad
R_2=\frac{1}{2}(Q_2+\sqrt{3}Q_{10}),\qquad
R_3=\frac{1}{2}(Q_3+\sqrt{3}Q_{8}),\nonumber \\
R_4&=&Q_4,\quad R_5=Q_5,\qquad R_6=Q_{12},\quad
R_7=\frac{1}{2}(Q_7+\sqrt{3}Q{14}),\nonumber \\
R_8&=&\frac{1}{2}(\sqrt{3}Q_{13}-Q_6).
\end{eqnarray}
The first three generators are those of $SO(3)_{\textrm{diag}}$. The
coset representative is given by
\begin{equation}
L=e^{a_1\sqrt{3}Y_5}e^{a_2\sqrt{3}(Y_7+\sqrt{3}Y_3)}e^{a_3\sqrt{3}(Y_8-\sqrt{3}Y_4)}\,
.
\end{equation}
The embedding tensor is similar to that of the $SU(2,1)$, but the
Catan-Killing is now $\eta_{SL(3)}$
\begin{equation}
\Theta=g\eta_{SL(3)}\, .
\end{equation}
The resulting potential is given by
\begin{eqnarray}
V&=&\frac{1}{64} g^2 \left[-6310-1848 \cosh(2 a_1)-66 \cosh(4 a_1)-6
\cosh(4 a_1-6 a_2)\right.\nonumber \\
& & -24 \cosh[2 (a_1-4 a_2)]+36 \cosh[2 (a_1-2 a_2)]-6 \cosh[4
(a_1-2 a_2)]\nonumber \\
& & +18 \cosh(4 a_1-2 a_2)+912 \cosh[2 (a_1-a_2)]-21 \cosh[4
(a_1-a_2)]\nonumber \\
& & -1860 \cosh(2 a_2)-30 \cosh(4 a_2)+12 \cosh(6 a_2)-36
\cosh(8 a_2)\nonumber \\
& & +912 \cosh[2 (a_1+a_2)]-21 \cosh[4 (a_1+a_2)]+18 \cosh[2 (2
a_1+a_2)]\nonumber \\
& & +36 \cosh[2 (a_1+2 a_2)]-6 \cosh[4 (a_1+2 a_2)]-24 \cosh[2
(a_1+4 a_2)]\nonumber \\
& & -6 \cosh(4 a_1+6 a_2)-48 \cosh^2 a_2 \left(221+80 \cosh(2
a_1)+3 \cosh(4 a_1)\right.\nonumber \\
& & -(137+156 \cosh(2 a_1)+3 \cosh(4 a_1)) \cosh(2 a_2)+4
\cosh^2 a_1 \times \nonumber \\
& &\left.(-5+\cosh(2 a_1)) \cosh(4 a_2)+8 \cosh^4 a_1 \cosh(6
a_2)\right) \cosh(2 a_3)\nonumber \\
& & +96 \cosh^2a_1 \left(3+\cosh(2 a_1)-4 \cosh^2a_1
\cosh(2 a_2)\right) \times \nonumber \\
& & (\cosh a_2+\cosh(3 a_2))^2 \cosh(6 a_3)-384 \cosh^4a_1
\cosh^4(2 a_2) \cosh(8 a_3)\nonumber \\
& & +6 \cosh(4 a_3) \left(290+320 \cosh(2 a_1)+6 \cosh(4
a_1)+\left(325+340 \cosh(2 a_1) \right. \right.\nonumber \\
& &\left. +7 \cosh(4 a_1)\right) \cosh(4 a_2)-8 \cosh^4a_1 \cosh(8
a_2)-8 (3+7 \cosh(2 a_1))\times \nonumber \\
& &\left. \left.\cosh(2 a_2) \sinh^2a_1-12 \cosh(6 a_2) \sinh^2(2
a_1)\right)\right].
\end{eqnarray}
We find some critical points shown below.
\begin{center}
\begin{tabular}{|c|c|c|c|c|}
  \hline
  critical point & $(a_1,a_2,a_3)$ & residual  & residual & $V_0$
  \\
  & & supersymmetry & gauge symmetry &
  \\ \hline
  1 & $(0,0,0)$ & $(4,0)$ & $SO(3)$ & $-16g^2$ \\ \hline
  2 & $(\cosh^{-1}\sqrt{3},0,0)$ & - & $SO(2)$ & $176g^2$ \\ \hline
  3 & $\left(0,\cosh^{-1}\sqrt{\frac{1+\sqrt{3}}{2}},0\right)$  & - & $SO(2)$ & $176g^2$
  \\
   & or $\left(0,0,\cosh^{-1}\sqrt{\frac{1+\sqrt{3}}{2}}\right)$ & & &
   \\ \hline
  4 & $(a_0,a_0,0)$ & - & - & $176g^2$ \\
  \hline
\end{tabular}
\end{center}
The $a_0$ is given by
$\cosh^{-1}\left(\frac{1}{2}\sqrt{\frac{3+[3(171-2\sqrt{7053})]^{\frac{1}{3}}+[3(171+2\sqrt{7053})]^{\frac{1}{3}}}{3}}\right)$.

\section{Some gaugings in $F_{4(4)}/USp(6)\times SU(2)$ coset
manifold}\label{F4} In this symmetric space, we have $F_{4(4)}$, the
split form of $F_4$, as the global symmetry while the representation
$R_0$ is $\mathbf{1053}$. The embedding tensor lives in the
$\mathbf{1}+\mathbf{324}$ representation. This case is similar to
the $N=9$ theory in which the coset space is given by
$F_{4(-20)}/SO(9)$ studied in \cite{AP}. The following subgroups of
$F_4$ can be gauged: $SO(9)$, $G_2\times SU(2)$, $USp(6)\times
SU(2)$ and $SU(3)\times SU(3)$. It has been shown in \cite{dewit}
that $SO(p)\times SO(9-p)\subset SO(9)$, $p=9,8,7,6,5$ can also be
gauged. Therefore, all real forms of the above subgroups that can be
embedded in $F_{4(4)}$ are admissible. These are given by:
\begin{itemize}
  \item $SO(5,4)$, $SO(5,3)$, $SO(4,4)$, $SO(5,2)\times SO(2)$, $SO(4,3)\times
  SO(2)$,\\
  $SO(5,1)\times SO(3)$, $SO(4,2)\times SO(3)$, $SO(4,1)\times
  SO(4)$,\\ and $SO(5)\times SO(4)$ or
  \begin{eqnarray}
  & & SO(5,p)\times SO(4-p),\qquad p=0,1,2,3,4,\nonumber \\
  & & SO(4,p)\times SO(5-p),\qquad p=1,2,3,4
  \end{eqnarray}
  \item $G_{2(2)}\times SL(2,\mathbb{R})$
  \item $USp(6)\times SU(2)$ and $Sp(6,\mathbb{R})\times SL(2,\mathbb{R})$
  \item $SU(3)\times SU(2,1)$
\end{itemize}
The maximal compact subgroup $SO(5)\times SO(4)\subset SO(5,4)$ is
embedded in the local symmetry $H$ as $USp(4)\times SU(2)\times
SU(2)\sim SO(5)\times SO(4)\subset USp(6)\times SU(2)$. The other
$SO$ type gauge groups can be embedded in $SO(5,4)$. The gauge group
$G_{2(2)}\times SL(2,\mathbb{R})$ can be considered as follow:
$G_{2(2)}$ is embedded in $SO(4,3)\subset SO(4,3)\times SO(2)$ while
the $SO(2)$ forms the compact subgroup of the $SL(2,\mathbb{R})$.
For $SU(3)\times SU(2,1)$, we first consider the maximal compact
subgroup $SU(3)\times SU(2)\times U(1)$. The $U(3)\sim SU(3)\times
U(1)$ is embedded in $U(3)\subset USp(6)$, and the $SU(2)$ is the
one appearing in the group $H=USp(6)\times SU(2)$.
\\
\indent Since the computation of the resulting scalar potentials is
more complicated in this case, we refrain from giving these
potentials in this work. The interested reader can do this
computation by using the information about the structure of
$F_{4(4)}/USp(6)\times SU(2)$ coset space given in the next
subsection and in the appendix.
\\
\indent It has been known that the ungauged $N=4$ theory with scalar
manifold $F_{4(4)}/USp(6)\times SU(2)$ can be obtained from the
dimensional reduction on $T^3$ of $N=(1,0)$ six dimensional
supergravity coupled to two tensor and two vector multiplets, see
\cite{Pope_3D_coset} for details. So, we expect to find $SO(4)$
$N=4$ Yang-Mills gauged supergravity in three dimensions from a
dimensional reduction of this six dimensional theory on $S^3$ with
$SO(4)$ being the isometry group of the $S^3$. As shown in
\cite{csym}, the resulting three dimensional theory is related to
the Chern-Simons gauged supergravity considered in this work with a
non-semisimple gauge group of the form $SO(4)\ltimes \mathbf{T}^6$.
The translational generators $\mathbf{T}^6$ transform as an adjoint
representation of the $SO(4)$.
\subsection{$SO(4)\ltimes \mathbf{T}^6$ gauging}
We now construct $N=4$ $SO(4)\ltimes \mathbf{T}^6$ gauged
supergravity with scalar manifold $F_{4(4)}/USp(6)\times SU(2)$.
According to \cite{max_ideal}, the split form $F_{4(4)}$ has nine
commuting generators, so it is possible to construct the gauge group
$SO(4)\ltimes \mathbf{T}^6\subset F_{4(4)}$. There is a systematic
procedure to find non-semisimple gaugings by boosting the admissible
semisimple ones, see \cite{nicolai3} for details. In this work, we
will directly look for this gauging by solving the consistency
constraints.
\\
\indent We begin with the identification of the gauge group
$SO(4)\ltimes \mathbf{T}^6$. An easy way to do this is to consider
the embedding of this group inside the $SO(5,4)\subset F_{4(4)}$. We
recall the explicit matrix form of the 52 generators of $F_4$ from
\cite{F4}. These generators are denoted by $c_i$'s in \cite{F4}. We
choose the $SO(9)$ subgroup by taking the corresponding generators
to be $c_i$ for $i=1,\ldots, 21, 30,\ldots, 36, 45,\ldots, 52$. It
is more convenient to relabel these generators in the form
$X^{ij}=X^{[ij]}$, $i,j =1,\ldots ,9$. The non-compact form
$SO(5,4)$ can be obtained by the Weyl unitarity trick namely
multiplying $X^{ij}$ for $i=1,\ldots , 5$ and $j=6,\ldots , 9$ by a
factor of $i$.
\\
\indent The relation between the $c_i$'s and the $X^{ij}$ has been
given in the appendix of \cite{AP}, and we refer the reader to this
work for the explicit form of $X^{ij}$'s. We now give the embedding
of $SO(4)\ltimes \mathbf{T}^6$ in $SO(5,4)\subset F_{4(4)}$. The
semisimple part $SO(4)$ is given by the diagonal subgroup of the
compact subgroup of $SO(4,5)$, $SO(4)=(SO(4)\times
SO(4))_{\textrm{diag}}$ with one of the $SO(4)$ being a subgroup of
$SO(5)$ in $SO(5)\times SO(4)\subset SO(4,5)$. The corresponding
generators are given by
\begin{equation}
J^{ab}=X^{ab}+X^{\hat{a}\hat{b}}, \qquad a,b=1,2,3,4\qquad
\textrm{and}\qquad \hat{a},\hat{b}=6,7,8,9\, .
\end{equation}
The generators of the translational part $\mathbf{T}^6$ are given by
\begin{equation}
t^{ab}=X^{ab}-X^{\hat{a}\hat{b}}+i(X^{a\hat{b}}+X^{\hat{a}b}).
\end{equation}
Notice the factor of $i$ indicating the non-compact generators of
$F_{4(4)}$. Using the $SO(9)$ algebra satisfied by $X^{ij}$, it can
be easily verified that $J^{ab}=J^{[ab]}$ and $t^{ab}=t^{[ab]}$
satisfy the algebra
\begin{equation}
\left[J^{ab},J^{cd}\right]=-4\delta^{[a[c}J^{d]b]},\qquad
\left[J^{ab},t^{cd}\right]=-4\delta^{[a[c}t^{d]b]},\qquad
\left[t^{ab},t^{cd}\right]=0\, .
\end{equation}
This algebra shows that the $t^{ab}$ indeed transform as the adjoint
representation of $SO(4)$ generated by $J^{ab}$. Therefore, $J^{ab}$
and $t^{ab}$ generate the non-semisimple group $SO(4)\ltimes
\mathbf{T}^6$.
\\
\indent Follow \cite{henning_AdS3_S3} in which $SO(4)\ltimes
\mathbf{T}^6$ $N=8$ gauged supergravity described the $S^3$
reduction of $N=(2,0)$ supergravity in six dimensions has been
constructed, we now decompose the generators $J^{ab}$ and $t^{ab}$
in terms of the self-dual and anti-self-dual parts
\begin{eqnarray}
J^{ab}_\pm &=&J^{ab}\pm\frac{1}{2}\epsilon_{abcd}J^{cd},\nonumber \\
t^{ab}_\pm &=&t^{ab}\pm\frac{1}{2}\epsilon_{abcd}t^{cd}\, .
\end{eqnarray}
These generate $(SO(3)^+\ltimes \mathbf{T}^3)\times (SO(3)^-\ltimes
\mathbf{T}^3) \sim SO(4)\ltimes \mathbf{T}^6$. As a general result
of \cite{csym}, the corresponding embedding tensor is of the form
\begin{equation}
\Theta = g_1\Theta_{\mc{AB}}+g_2\Theta_{\mc{BB}}
\end{equation}
where $\mc{A}$ and $\mc{B}$ describe the semisimple and
translational parts, respectively. In the self-dual and
anti-self-dual basis, the ansatz for the embedding is given by
\begin{eqnarray}
\Theta &=&
g_{1+}(\Theta_{\mc{A}^+\mc{B}^+}+\Theta_{\mc{B}^+\mc{A}^+})+g_{1-}(\Theta_{\mc{A}^-\mc{B}^-}+\Theta_{\mc{B}^-\mc{A}^-})\nonumber
\\ &
&+g_{2+}\Theta_{\mc{B}^+\mc{B}^+}+g_{2-}\Theta_{\mc{B}^-\mc{B}^-}\,
.
\end{eqnarray}
The quadratic and linear constraints impose the conditions
\begin{equation}
g_{2-}=-g_{2+}=g_2,\qquad \textrm{and}\qquad g_{1-}=-g_{1+}=g_1\, .
\end{equation}
This is similar to the $N=8$ theory considered in
\cite{henning_AdS3_S3} in which the consistency conditions also
require the relative minus sign between the self-dual and
anti-self-dual parts.
\\
\indent We then end up with the embedding tensor of the form
\begin{eqnarray}
\Theta
&=&g_1(\Theta_{\mc{A}^+\mc{B}^+}-\Theta_{\mc{A}^-\mc{B}^-}+\Theta_{\mc{B}^+\mc{A}^+}-\Theta_{\mc{B}^-\mc{A}^-})\nonumber
\\
& &+g_{2}(\Theta_{\mc{B}^+\mc{B}^+}-\Theta_{\mc{B}^-\mc{B}^-}) .
\end{eqnarray}
\indent Furthermore, as we will see, it turns out that the existence
of the maximal supersymmetric $AdS_3$ vacua at $L=\mathbf{I}$
requires the relation $g_2=g_1$. This is again much similar to the
$N=8$ theory studied in \cite{henning_AdS3_S3}.
\\
\indent Since this theory describes an $S^3$ reduction of the
$(1,0)$ six dimensional supergravity coupled to two vector and two
tensor multiplets, it is interesting to further study the associated
scalar potential. This is certainly useful in the AdS/CFT
correspondence. We again refer the reader to the needed formulae in
the appendix. There are 28 scalars in the coset manifold
$F_{4(4)}/USp(6)\times SU(2)$. With the 26 dimensional fundamental
representation of $F_4$, it is extremely difficult to study the full
scalar potential on the 28 dimensional scalar manifold. We then
apply the group theory argument of \cite{warner} to compute the
scalar potential on a submanifold of the full 28-dimensional scalar
manifold which is invariant under a certain subgroup of the gauge
group.
\\
\indent There are two scalars invariant under the $SO(4)$ part of
the gauge group. The coset representative parametrized by these two
singlets is given by
\begin{equation}
L=e^{b_1\bar{Y}_1}e^{b_2\bar{Y}_2}
\end{equation}
where
\begin{equation}
\bar{Y}_1=Y_3-Y_4-Y_5-Y_8,\qquad
\bar{Y}_2=Y_9+Y_{11}-Y_{15}-Y_{17}+Y_{19}-Y_{21}-Y_{25}+Y_{27}\, .
\end{equation}
We finally obtain the scalar potential
\begin{eqnarray}
V&=&256\left[2\cosh(\sqrt{2}b_1)\cosh
(2b_2)-2\sinh(2b_2)\right]^2\left[5g_1^2-2g_2^2 \right. \nonumber \\
& &\left.
+2g_2\left(2g_2\cosh^2(2\sqrt{2}b_2)+3g_2\cosh(4b_2)+7g_1\sinh(2b_2)\right.
\right.  \nonumber \\
& & \left. \left.
-\cosh(\sqrt{2}b_1)\cosh(2b_2)(7g_1+8g_2\sinh(2b_2)) \right)
\right].
\end{eqnarray}
From this potential, we find that the potential admits a critical
point at $b_1=b_2=0$ or $L=\mathbf{I}$ only for
\begin{equation}
g_2=g_1,\qquad g_2=\frac{5}{16}g_1\, .
\end{equation}
The first one corresponds to the maximal supersymmetric $AdS_3$
point with $V_0=-1024g_1^2$. This can be checked by the condition
given in \cite{dewit}
\begin{equation}
A_1^{IK}A_1^{KJ}\epsilon^J=-\frac{V_0}{4}\epsilon^I
\end{equation}
which states that the unbroken supersymmetries are given by the
eigenvalues of the $A_1$ tensor that satisfy the above condition at
the critical point. It can be verified that only $g_2=g_1$ satisfies
this condition.
\\
\indent The second possibility gives $V_0=1440g_1^2$ which is of a
dS type. Therefore, this supergravity theory admits both dS and AdS
vacua at $L=\mathbf{I}$ depending on the value of $g_2$. Our main
interest here is for the $g_2=g_1$ case, so we will further explore
this case.
\\
\indent Setting $g_2=g_1=g$, we find the potential
\begin{eqnarray}
V&=&1024g^2\left[\sinh(2b_2)-\cosh(\sqrt{2}b_1)\cosh(2b_2)\right]^2\left[3+4\cosh(2\sqrt{2}b_1)\cosh^2(2b_2)\right.\nonumber
\\
&
&\left.+6\cosh(4b_2)-2\cosh(\sqrt{2}b_1)(7\cosh(2b_2)+4\sinh(4b_2))
\right].
\end{eqnarray}
Analyzing this potential gives the following non-trivial critical
points.
\begin{itemize}
  \item At $b_1=0$ and $b_2=\frac{1}{2}\ln \frac{16}{5}$, there is a
  dS critical point with $V_0=\frac{1125}{8}g^2$.
  \item There is a class of Minkowski vacua for
  $\cosh(\sqrt{2}b_1)=\tanh(2b_2)$ with $V_0=0$ and $N=4$ supersymmetry.
\end{itemize}
We can consider a smaller residual symmetry namely $SO(3)$ subgroup
of $SO(4)$ generated by $J^{12}$, $J^{13}$ and $J^{23}$. There are
five singlets given by
\begin{eqnarray}
\tilde{Y}_1&=&Y_3-Y_5,\qquad \tilde{Y}_2=Y_{17}-Y_{27},\qquad
\tilde{Y}_3=Y_{18}-Y_{28},\nonumber \\
\tilde{Y}_4&=&Y_4+Y_8,\qquad
\tilde{Y}_5=Y_9+Y_{11}-Y_{15}+Y_{19}-Y_{21}-Y_{25}\, .
\end{eqnarray}
Unfortunately, the computation of the potential turns out to be
extremely difficult. Therefore, we will leave this for future works.
\\
\indent There is another subgroup of $SO(5,4)$ which is of the form
$SO(4)\ltimes (\mathbf{T}^6,\hat{\mathbf{T}}^4)$. The
$\hat{\mathbf{T}}^4$ transform as a vector ($\mathbf{4}$) of $SO(4)$
and close onto the translational symmetry $\mathbf{T}^6$. According
to \cite{csym}, the theory with this gauge group is on-shell
equivalent to the Yang-Mills gauged supergravity with the gauge
group $SO(4)$ coupled to four massive vector fields corresponding to
the nilpotent generators of $\hat{\mathbf{T}}^4$. The generators of
this group are those of the $SO(4)\ltimes \mathbf{T}^6$ together
with the four generators of $\hat{\mathbf{T}}^4$ given by
\begin{equation}
\hat{T}^a=X^{a,5}-iX^{5,a+5},\qquad a=1,2,3,4\, .
\end{equation}
All of these generator satisfy the above mentioned algebra as can be
readily verified.
\\
\indent The embedding tensor for this gauge group is that of the
$SO(4)\ltimes \mathbf{T}^6$ with an additional component of the form
$g_3\Theta_{\hat{\mathbf{T}}\hat{\mathbf{T}}}$. It turns out that
consistency conditions require $g_3=0$. So, the whole $SO(4)\ltimes
(\mathbf{T}^6,\hat{\mathbf{T}}^4)$ cannot be gauged.

\section{Some gaugings in $E_{6(2)}/SU(6)\times SU(2)$, $E_{7(-5)}/SO(12)\times SU(2)$ and $E_{8(-24)}/E_7\times SU(2)$ coset
manifolds}\label{E678} In this section, we consider some admissible
gauge groups obtained by applying the general group theory argument
presented in section \ref{N4theory}. The global symmetry $G$ in
these cases is large, and it is even more difficult than the
$F_{4(4)}/USp(6)\times SU(2)$ case to give all admissible gauge
groups. Therefore, we will not attempt to give an exhaustive list
for these target manifolds in this work but simply provide some
examples.
\subsection{Examples of gaugings in $E_{6(2)}/SU(6)\times SU(2)$ coset manifold}
In this case, the group $G$ and representation $R_0$ are $E_{6(2)}$
and $\mathbf{2430}$. The corresponding embedding tensor lives in the
representation $\mathbf{1}+\mathbf{650}$. The group theory structure
is similar to the $N=10$ theory whose scalar potentials have been
studied in \cite{AP2}, but the scalar manifold is described by
$E_{6(-14)}/SO(10)\times U(1)$ coset space. The suitable real forms
of the $D_5\times U(1)$, $A_5\times A_1$, $F_4$ and $G_2\times A_2$
subgroups can be gauged.
\\
\indent Some admissible gauge groups are given by:
\begin{itemize}
  \item $SO(6,4)\times U(1)$, $SO(6,3)\times U(1)$, $SO(6,2)\times SO(2)\times
  U(1)$,\\
  $SO(4,4)\times SO(2)\times U(1)$, $SO(6,1)\times SO(3)\times U(1)$, $SO(4,3)\times SO(3)\times
  U(1)$,\\
  $SO(6)\times SO(4)\times U(1)$, $SO(4,2)\times SO(4)\times U(1)$,\\ and $SO(4,1)\times
  SO(5)$ or
  \begin{eqnarray}
  & & SO(6,p)\times SO(4-p)\times U(1),\qquad p=0,1,2,3,4,\nonumber \\
  & & SO(4,p)\times SO(6-p)\times U(1)^{1-\delta_{1p}},\qquad p=1,2,3,4
  \end{eqnarray}
  \item $SU(6)\times SU(2)$ and $SU(3,3)\times SL(2,\mathbb{R})$
  \item $F_{4(4)}$
  \item $G_{2(2)}\times SU(2,1)$\, .
\end{itemize}
The embedding of the $SO(6,4)\times U(1)$ can be given as follow. We
first consider the decomposition of $SU(6)$ to $SU(4)\times
SU(2)\times U(1)$. Together with the additional $SU(2)$ factor from
the group $H$ and using $SU(4)\sim SO(6)$ and $SU(2)\times SU(2)\sim
SO(4)$, we find the maximal compact subgroup $SO(6)\times
SO(4)\times U(1)\subset SO(6,4)\times U(1)$. The other real forms
can be embedded in $SO(6,4)\times U(1)$. The $USp(6)\subset SU(6)$
together with the $SU(2)$ form the maximal compact subgroup of
$F_{4(4)}$. Finally, $G_{2(2)}\times SU(2,1)$ can be embedded in
$SO(4,3)\times SO(3)\times U(1)$ with $G_{2(2)}\subset SO(4,3)$ and
the $SO(3)\times U(1)$ being the maximal compact subgroup of
$SU(2,1)$.

\subsection{Examples of gaugings in $E_{7(-5)}/SO(12)\times SU(2)$ coset manifold}
With this target manifold, the group $G$ and representation $R_0$
are given by $E_{7(-5)}$ and $\mathbf{7371}$. The corresponding
embedding tensor lives in the representation
$\mathbf{1}+\mathbf{1539}$. This case is the same as $N=12$ theory
in which the scalar manifold is uniquely determined to be
$E_{7(-5)}/SO(12)\times SU(2)$ \cite{dewit1}. Some admissible gauge
groups have already been given in \cite{dewit}. We simply repeat
them here for completeness.
\\
\indent They are given by:
\begin{itemize}
  \item $SO(p)\times SO(12-p)\times U(1)$ for $p=0,\ldots 5$, $SO(6)\times
  SO(6)$, \\
  and $SO^*(12)\times SL(2,\mathbb{R})$
  \item $E_{6(2)}\times U(1)$
  \item $F_{4(-20)}\times SU(2)$
  \item $G_{2(2)}\times USp(6)$\, .
\end{itemize}

\subsection{Examples of gaugings in $E_{8(-24)}/E_7\times SU(2)$ coset manifold}
In this case, the group $G$ and representation $R_0$ are given by
$E_{8(-24)}$ and $\mathbf{27000}$. The corresponding embedding
tensor lives in the representation $\mathbf{1}+\mathbf{3875}$. This
case is similar to the maximal $N=16$ theory but with different real
form of $E_8$ namely the group $G$ is $E_{8(8)}$ for the maximal
case. The study of scalar potentials for some semisimple gauge
groups has been given in \cite{N16Vacua}. The suitable real forms of
the following subgroups can be gauged: $D_4\times D_4$, $G_2\times
F_4$, $E_6\times A_2$ and $E_7\times A_1$.
\\
\indent Some admissible gauge groups are given by:
\begin{itemize}
  \item $SO(4,4)\times SO(8)$, $SO(7,1)\times SO(5,3)$ and $SO(6,2)\times
  SO(6,2)$ \\
  or
  \begin{displaymath}
  SO(8-p,p)\times SO(4+p,4-p),\qquad p=0,1,2
  \end{displaymath}
  \item $F_4\times G_{2(2)}$, $F_{4(4)}\times G_2$ and $F_{4(-20)}\times G_{2(2)}$
  \item $E_{6}\times SU(2,1)$, $E_{6(2)}\times SU(2,1)$ and $E_{6(-26)}\times SL(3,\mathbb{R})$
  \item $E_7\times SU(2)$ and $E_{7(-25)}\times SL(2,\mathbb{R})$\, .
\end{itemize}
The $SO(4,4)\times SO(8)$ and other real forms can be embedded in
the maximal subgroup $SO(12,4)\subset E_{8(-24)}$. Using the
decompositions $E_7\times SU(2)\rightarrow F_4\times SU(2)\times
SU(2)$ and $E_7\times SU(2)\rightarrow G_2\times USp(6)\times
SU(2)$, we immediately see the embedding of $F_4\times G_{2(2)}$ and
$F_{4(4)}\times G_2$, respectively. The embedding of the real form
$F_{4(-20)}\times G_{2(2)}$ can be seen as follow. We first
decompose $SO(12,4)\rightarrow SO(9)\times SO(3,4)$. The $SO(9)$
becomes the maximal compact subgroup of $F_{4(-20)}$, and $G_{2(2)}$
is embedded in $SO(4,3)$. Using the embedding of $E_6\times
U(1)\subset E_7$, we can see the embedding of $E_{6}\times SU(2,1)$
while the decomposition $E_7\rightarrow SO(12)\times
SU(2)\rightarrow U(6)\times SU(2)$ gives the embedding of
$E_{6(2)}\times SU(2,1)$. Finally, $E_{6(-26)}\times
SL(3,\mathbb{R})$ is embedded in $E_{8(-24)}$ via the decomposition
$E_7\times SU(2)\rightarrow F_4\times SU(2)\times SU(2)$ with the
maximal compact subgroup $SO(3)$ of $SL(3,\mathbb{R})$ being
$SU(2)_{\textrm{diag}}$. The 112 non-compact generators of
$E_{8(-24)}$ transform as $(\mathbf{56},\mathbf{2})$ under
$E_7\times SU(2)$ which is further decomposed into
$(\mathbf{26},\mathbf{2},\mathbf{2})+(\mathbf{1},\mathbf{4},\mathbf{2})$
under $F_4\times SU(2)\times SU(2)$. Under $F_4\times
SU(2)_{\textrm{diag}}$, they transform as
\begin{displaymath}
(\mathbf{56},\mathbf{2})\rightarrow
(\mathbf{26},\mathbf{1})+(\mathbf{26},\mathbf{3})+(\mathbf{1},\mathbf{3})+(\mathbf{1},\mathbf{5}).
\end{displaymath}
The $(\mathbf{26},\mathbf{1})$ enlarges the $F_4$ to $E_{6(-26)}$
while the $(\mathbf{1},\mathbf{5})$ becomes five non-compact
generators of $SL(3,\mathbb{R})$.

\section{Conclusions}\label{conclusions}
In this paper, we have studied gaugings of $N=4$ gauged supergravity
in three dimensions. The scalar target spaces considered here are in
the form of the exceptional coset spaces. In the $G_{2(2)}/SO(4)$
case, we have listed all admissible gauge groups as well as study
their scalar potentials and some of the corresponding critical
points. We have pointed out that the $SO(3)\ltimes \mathbf{T}^3$
cannot be gauged since this gauge group cannot be embedded in the
$G_{2(2)}$. This immediately leads to a puzzle. The ungauged version
of this theory can be obtained from $T^2$ reduction of the minimal
($N=2$) supergravity in five dimensions. It has been proposed in
\cite{csym} that the three dimensional gauged theory with scalar
manifold $G_{2(2)}/SO(4)$ and gauge group $SO(3)\ltimes
\mathbf{T}^3$ would describe the $N=2$ five dimensional supergravity
reduced on an $S^2$. The spectrum of the Kaluza-Klein reduction has
been studied in \cite{AdS3S2_spectra1} and \cite{AdS3S2_spectra2}
together with its dual SCFT. On the other hand, it has been pointed
out in \cite{Pope_sphere} that although the symmetry of the $T^2$
reduction of the minimal supergravity in five dimensions get
enhanced to $G_{2(2)}$, there does not seem to be a possibility of a
consistent $S^2$ reduction. This is precisely in agreement with what
has been found here. It could be that the consistent $S^2$ reduction
at the full non-linear level may not exist. It is interesting to
find the three dimensional description of this reduced theory if the
reduction can be achieved.
\\
\indent For the theory with $F_{4(4)}/USp(6)\times SU(2)$, we have
identified some gauge groups which are maximal subgroups of
$F_{4(4)}$. As one of the main results of this paper, we have
constructed an $N=4$ theory with gauge group $SO(4)\ltimes
\mathbf{T}^6$. The resulting theory is on-shell equivalent to
Yang-Mills gauged supergravity with $SO(4)$ gauge group according to
the general result of \cite{csym}. So, we expect that the
$SO(4)\ltimes \mathbf{T}^6$ gauged supergravity with
$F_{4(4)}/USp(6)\times SU(2)$ scalar manifold can be obtained from
the $S^3$ reduction of the $(1,0)$ six dimensional supergravity
coupled to two vector and two tensor multiplets. In this case, it is
also interesting to study its explicit reduction in the same way as
the reduction on the $SU(2)$ group manifold studied in \cite{KK},
\cite{PopeSU2} and \cite{PopeSU22}. From the general results of
\cite{Pope_sphere}, there do not seem to be any obstacles in this
case. Furthermore, we have shown that the theory constructed here
admits both AdS and dS vacua at $L=\mathbf{I}$ depending on the
value of the coupling constants. The situation is very similar to
$N=4$ gauged supergravity in four dimensions with $SU(2)\times
SU(2)$ gauge group in which the relative values of the two couplings
determine whether the vacuum is a supersymmetric $AdS_4$ or a
non-supersymmetric $dS_4$ solution, see for example the discussion
in \cite{Hull_DW}.
\\
\indent We end the paper by considering some examples of admissible
gauge groups in the case of $E_{6(2)}/SU(6)\times SU(2)$,
$E_{7(-5)}/SO(12)\times SU(2)$ and $E_{8(-24)}/E_7\times SU(2)$
scalar manifolds. All of the gauge groups presented here are maximal
subgroups of the corresponding global symmetry $G$. The detailed
study is needed in order to find a larger class of admissible gauge
groups for these theories. It is of interest to study their
non-semisimple gaugings which might give some insights to the higher
dimensional origin of these theories.
\\
\indent It could also be interesting to study the scalar potentials
as well as the corresponding critical points for the gauge groups
identified here. This is useful in the study of holographic RG flows
describing the deformations of the dual SCFT's. We hope to come back
to these issues in the future works.
\acknowledgments The author would like to thank the Abdus Salam
International Centre for Theoretical Physics (ICTP) for hospitality
and computing facilities as well as International School for
Advanced Studies (SISSA) for the support when the early stage of
this work was initiated. He also gratefully thanks Henning Samtleben
for invaluable correspondences. This work is partially supported by
Thailand Center of Excellence in Physics through the
ThEP/CU/2-RE3/11 project and Chulalongkorn University through
Ratchadapisek Sompote Endowment Fund.
\appendix
\section{Essential formulae}\label{detail}
In this appendix, we collect some useful formulae for three
dimensional gauged supergravity with symmetric scalar target
manifolds. We also give some details about the explicit construction
of the coset space $G_{2(2)}/SO(4)$ and $F_{4(4)}/USp(6)\times
SU(2)$ which are used in the main text and might be useful for
further investigations.
\\
\indent We begin with the formulae for a symmetric space of the form
$G/H$ in which $G$ and $H$ are the global and local symmetry groups,
respectively. The $G$ algebra is given by
\begin{eqnarray}
\left[T^{IJ},T^{KL}\right]&=&-4\delta^{[I[K}T^{L]J]}, \qquad
\left[T^{IJ},Y^A\right]=-\frac{1}{2}f^{IJ,AB}Y_B, \nonumber \\
\left[X^\alpha,X^\beta\right]&=&f^{\alpha
\beta}_{\phantom{as}\gamma}X^\gamma,\qquad
\left[X^\alpha,Y^A\right]=h^{\alpha
\phantom{a}A}_{\phantom{a}B}Y^B, \nonumber \\
\left[Y^{A},Y^{B}\right]&=&\frac{1}{4}f^{AB}_{IJ}T^{IJ}+\frac{1}{8}C_{\alpha\beta}h^{\beta
AB}X^\alpha \label{Galgebra}
\end{eqnarray}
$T^{IJ}$'s and $X^\alpha$'s generate $SO(N)\times H'$, and $Y^A$'s
are non-compact generators transforming in a spinor of $SO(N)$. We
refer the reader to \cite{dewit} for other notations. The coset
representative $L$ transforming under $G$ and $H$ by left and right
multiplications can be used to define the map $\mc{V}$ by the
relation
\begin{equation}
L^{-1}t^\mathcal{M}L=\frac{1}{2}\mathcal{V}^{\mathcal{M}}_{\phantom{as}IJ}T^{IJ}+\mathcal{V}^\mathcal{M}_{\phantom{as}\alpha}X^\alpha+
\mathcal{V}^\mathcal{M}_{\phantom{as}A}Y^A\, .\label{cosetFormula}
\end{equation}
The metric on the target space $g_{ij}$ can be computed from the
vielbein $e^A_i$ which is in turn encoded in
\begin{equation}
L^{-1} \partial_i L= \frac{1}{2}Q^{IJ}_i T^{IJ}+Q^\alpha_i
X^{\alpha}+e^A_i Y^A,\label{cosetFormula1}
\end{equation}
where $Q_i^{IJ}$ and $Q^\alpha_i$ are the composite connections for
the $SO(N)$ and $H'$, respectively.
\\
\indent Given the map $\mc{V}$ from \eqref{cosetFormula}, the
T-tensor can be straightforwardly computed from the embedding tensor
by using \eqref{T_tensor_def}. In order to compute the scalar
potential, we need to construct the $A_1$ and $A_2$ tensors. They
are given in terms of the T-tensor components by \cite{dewit}
\begin{eqnarray}
A_1^{IJ}&=&-\frac{4}{N-2}T^{IM,JM}+\frac{2}{(N-1)(N-2)}\delta^{IJ}T^{MN,MN},\nonumber\\
A_{2j}^{IJ}&=&\frac{2}{N}T^{IJ}_{\phantom{as}j}+\frac{4}{N(N-2)}f^{M(I
m}_{\phantom{as}j}T^{J)M}_{\phantom{as}m}+\frac{2}{N(N-1)(N-2)}\delta^{IJ}f^{KL\phantom{a}m}_{\phantom{as}j}T^{KL}_{\phantom{as}m}\,
. \phantom{asad}
\end{eqnarray}
These two tensors together with the third one, $A_3$, appear in the
gauged Lagrangian as fermion mass-like terms \cite{dewit}. Finally,
the scalar potential can be computed by using
\begin{equation}
V=-\frac{4}{N}g^2\left(A_1^{IJ}A_1^{IJ}-\frac{1}{2}Ng^{ij}A_{2i}^{IJ}A_{2j}^{IJ}\right).
\end{equation}
\subsection{Useful formulae for $G_{2(2)}/SO(4)$ coset}
We give the explicit form of the various $\mc{V}$ maps and T-tensors
for all the gauge groups studied in the case of $G_{2(2)}/SO(4)$
coset manifold. These are relevant for computing the corresponding
scalar potentials. The repeated indices are summed over the given
values.
\begin{itemize}
  \item $SO(4)$ gauging:
  \begin{eqnarray}
  {\mc{V}_{(1),(2)}}^{i,IJ}&=&-\frac{1}{3}\textrm{Tr}(L^{-1}J^{(1),(2)}_iLT^{IJ}),\nonumber
  \\
  {\mc{V}_{(1),(2)}}^{i,A}&=&\frac{1}{3}\textrm{Tr}(L^{-1}J^{(1),(2)}_iLY^{A}),\qquad
  i=1,2,3,\nonumber \\
  T^{IJ,KL}&=&g\left({\mc{V}_{(2)}}^{i,IJ}{\mc{V}_{(2)}}^{i,KL}
  -3{\mc{V}_{(1)}}^{i,IJ}{\mc{V}_{(1)}}^{i,KL}\right),\nonumber \\
  T^{IJ,A}&=&g\left({\mc{V}_{(2)}}^{i,IJ}{\mc{V}_{(2)}}^{i,A}
  -3{\mc{V}_{(1)}}^{i,IJ}{\mc{V}_{(1)}}^{i,A}\right)
  \end{eqnarray}
  \item $SL(2,\mathbb{R})\times SL(2,\mathbb{R})$ gauging:
  \begin{eqnarray}
  {\mc{V}_{(1),(2)}}^{i,IJ}&=&-\frac{1}{3}\textrm{Tr}(L^{-1}j^{(1),(2)}_iLT^{IJ}),\nonumber
  \\
  {\mc{V}_{(1),(2)}}^{i,A}&=&\frac{1}{3}\textrm{Tr}(L^{-1}j^{(1),(2)}_iLY^{A}),\qquad
  i=1,2,3,\nonumber \\
  T^{IJ,KL}&=&g\left({\mc{V}_{(2)}}^{i,IJ}{\mc{V}_{(2)}}^{j,KL}
  -3{\mc{V}_{(1)}}^{i,IJ}{\mc{V}_{(1)}}^{j,KL}\right)\eta^{SL(2)}_{ij},\nonumber
  \\
  T^{IJ,A}&=&g\left({\mc{V}_{(2)}}^{i,IJ}{\mc{V}_{(2)}}^{j,A}
  -3{\mc{V}_{(1)}}^{i,IJ}{\mc{V}_{(1)}}^{j,A}\right)\eta^{SL(2)}_{ij}
  \end{eqnarray}
  \item $SU(2,1)$ gauging:
  \begin{eqnarray}
  \mc{V}^{aIJ}&=&-\frac{1}{3}\textrm{Tr}(L^{-1}Q_aLT^{IJ}),\qquad \mc{V}^{a,A}=\frac{1}{3}\textrm{Tr}(L^{-1}Q_aLY^{A}),
  \qquad a=1,\ldots, 8,\nonumber \\
  T^{IJ,KL}&=&g\mc{V}^{a,IJ}\mc{V}^{b,KL}\eta^{SU(2,1)}_{ab},\nonumber
  \\
  T^{IJ,A}&=&g\mc{V}^{a,IJ}\mc{V}^{b,A}\eta^{SU(2,1)}_{ab}
  \end{eqnarray}
  \item $SL(3,\mathbb{R})$ gauging:
  \begin{eqnarray}
  \mc{V}^{a,IJ}&=&-\frac{1}{3}\textrm{Tr}(L^{-1}R_aLT^{IJ}),\qquad \mc{V}^{a,A}=\frac{1}{3}\textrm{Tr}(L^{-1}R_aLY^{A}),
  \qquad a=1,\ldots, 8,\nonumber \\
  T^{IJ,KL}&=&g\mc{V}^{a,IJ}\mc{V}^{b,KL}\eta^{SL(3)}_{ab},\nonumber
  \\
  T^{IJ,A}&=&g\mc{V}^{a,IJ}\mc{V}^{b,A}\eta^{SL(3)}_{ab}
  \end{eqnarray}
\end{itemize}
The three almost complex structures are given by
\begin{eqnarray}
f_2&=&-\left(
      \begin{array}{cc}
        1 & 0 \\
        0 & 1 \\
      \end{array}
    \right)\otimes \left(
                     \begin{array}{cc}
                       0 & 1 \\
                       -1 & 0 \\
                     \end{array}
                   \right)\otimes \left(
      \begin{array}{cc}
        0 & 1 \\
        1 & 0 \\
      \end{array}
    \right),\nonumber \\
    f_3&=&
    -\left(
      \begin{array}{cc}
        1 & 0 \\
        0 & 1 \\
      \end{array}
    \right)\otimes \left(
                     \begin{array}{cc}
                       0 & 1 \\
                       -1 & 0 \\
                     \end{array}
                   \right)\otimes \left(
      \begin{array}{cc}
        1 & 0 \\
        0 & -1 \\
      \end{array}
    \right),\nonumber \\
    f_4&=&
    -\left(
      \begin{array}{cc}
        1 & 0 \\
        0 & 1 \\
      \end{array}
    \right)\otimes \left(
                     \begin{array}{cc}
                       1 & 0 \\
                       0 & 1 \\
                     \end{array}
                   \right)\otimes \left(
      \begin{array}{cc}
        0 & 1 \\
        -1 & 0 \\
      \end{array}
    \right).
\end{eqnarray}
The full tensor $f^{IJ}$ can be straightforwardly obtained by the
relation \eqref{fIJ}.

\subsection{Useful formulae for $F_{4(4)}/USp(6)\times SU(2)$ coset}
The R-Symmetry group $SU(2)\sim SO(3)$ is generated by
\begin{equation}
T^{12}=\frac{1}{2}(c_{52}-c_{21}),\qquad
T^{13}=-\frac{1}{2}(c_{51}+c_{35}),\qquad
T^{23}=-\frac{1}{2}(c_{36}-c_{50}).
\end{equation}
The generators for the group $H'=USp(6)$ are given by
\begin{eqnarray}
q_i&=&\frac{c_i}{\sqrt{2}},\qquad i=1,\ldots , 10, \nonumber \\
q_{11}&=&\frac{1}{2}(c_{21}+c_{52}),\qquad
q_{12}=\frac{1}{2}(c_{51}-c_{35}),\qquad
q_{13}=\frac{1}{2}(c_{50}+c_{36}),\nonumber \\
q_{14}&=&\frac{1}{2}(c_{22}+c_{38}),\qquad
q_{15}=\frac{1}{2}(c_{23}-c_{37}),\qquad
q_{16}=\frac{1}{2}(c_{24}+c_{41}),\nonumber \\
q_{17}&=&\frac{1}{2}(c_{25}+c_{44}),\qquad
q_{18}=\frac{1}{2}(c_{26}-c_{39}),\qquad
q_{19}=\frac{1}{2}(c_{27}-c_{43}),\nonumber \\
q_{20}&=&\frac{1}{2}(c_{28}+c_{42}),\qquad
q_{21}=\frac{1}{2}(c_{29}-c_{40}).
\end{eqnarray}
Finally, the 28 non-compact generators are
\begin{eqnarray}
Y_{1}&=&\frac{i}{2}(c_{22}-c_{38}),\qquad
Y_{2}=\frac{i}{2}(c_{23}+c_{37}),\qquad
Y_{3}=\frac{i}{2}(c_{24}-c_{41}),\nonumber \\
Y_{4}&=&\frac{i}{2}(c_{25}-c_{44}),\qquad
Y_{5}=\frac{i}{2}(c_{26}+c_{39}),\qquad
Y_{6}=\frac{i}{2}(c_{27}+c_{43}),\nonumber \\
Y_{7}&=&\frac{i}{2}(c_{28}-c_{42}),\qquad
Y_{8}=\frac{i}{2}(c_{29}+c_{40}),\qquad
Y_{9}=\frac{i}{2}(c_{11}+c_{30}),\nonumber \\
Y_{10}&=&\frac{i}{2}(c_{12}+c_{31}),\qquad
Y_{11}=\frac{i}{2}(c_{13}+c_{32}),\qquad
Y_{12}=\frac{i}{2}(c_{14}+c_{33}),\nonumber \\
Y_{13}&=&\frac{i}{2}(c_{15}+c_{34}),\qquad
Y_{14}=\frac{i}{2}(c_{16}+c_{45}),\qquad
Y_{15}=\frac{i}{2}(c_{17}+c_{46}),\nonumber \\
Y_{16}&=&\frac{i}{2}(c_{18}+c_{47}),\qquad
Y_{17}=\frac{i}{2}(c_{19}+c_{48}),\qquad
Y_{18}=\frac{i}{2}(c_{20}+c_{49}),\nonumber \\
Y_{19}&=&\frac{i}{2}(c_{11}-c_{30}),\qquad
Y_{20}=\frac{i}{2}(c_{12}-c_{31}),\qquad
Y_{21}=\frac{i}{2}(c_{13}-c_{32}),\nonumber \\
Y_{22}&=&\frac{i}{2}(c_{14}-c_{33}),\qquad
Y_{23}=\frac{i}{2}(c_{15}-c_{34}),\qquad
Y_{24}=\frac{i}{2}(c_{16}-c_{45}),\nonumber \\
Y_{25}&=&\frac{i}{2}(c_{17}-c_{46}),\qquad
Y_{26}=\frac{i}{2}(c_{18}-c_{47}),\qquad
Y_{27}=\frac{i}{2}(c_{19}-c_{48}),\nonumber \\
Y_{28}&=&\frac{i}{2}(c_{20}-c_{49}).
\end{eqnarray}
The $f^{IJ}$ tensors can be computed by using
$\left[T^{IJ},Y^A\right]$ in the $G$ algebra. With a suitable
normalization, they are given by
\begin{equation}
f^{IJ}_{AB}=-2\textrm{Tr}(Y^B\left[T^{IJ},Y^A\right]).
\end{equation}
We now give various components of the $\mc{V}$ map. They are
computed by
\begin{eqnarray}
\mc{{V_{\mc{A}}}}^{ab,IJ}&=&-\frac{1}{3}\textrm{Tr}(L^{-1}J^{ab}LT^{IJ}),\qquad
\mc{{V_{\mc{B}}}}^{ab,IJ}
=-\frac{1}{3}\textrm{Tr}(L^{-1}t^{ab}LT^{IJ}),\nonumber
\\
\mc{{V_{\mc{A}}}}^{ab,A}&=&\frac{1}{3}\textrm{Tr}(L^{-1}J^{ab}LY^{A}),
\qquad
\mc{{V_{\mc{B}}}}^{ab,A}=\frac{1}{3}\textrm{Tr}(L^{-1}t^{ab}LY^{A}).
\end{eqnarray}
The T-tensors are given by
\begin{eqnarray}
T^{IJ,KL}&=&g_1\left(\mc{{V_{\mc{A}}}}^{ab,IJ}\mc{{V_{\mc{B}}}}^{cd,KL}+
\mc{{V_{\mc{B}}}}^{ab,IJ}\mc{{V_{\mc{A}}}}^{cd,KL}\right)\epsilon_{abcd}\nonumber
\\
& &+
g_2\mc{{V_{\mc{B}}}}^{ab,IJ}\mc{{V_{\mc{B}}}}^{cd,KL}\epsilon_{abcd},\nonumber
\\
T^{IJ,A}&=&g_1\left(\mc{{V_{\mc{A}}}}^{ab,IJ}\mc{{V_{\mc{B}}}}^{cd,A}+
\mc{{V_{\mc{B}}}}^{ab,IJ}\mc{{V_{\mc{A}}}}^{cd,A}\right)\epsilon_{abcd}\nonumber
\\
& &+
g_2\mc{{V_{\mc{B}}}}^{ab,IJ}\mc{{V_{\mc{B}}}}^{cd,A}\epsilon_{abcd}\,
.
\end{eqnarray}
In the above equation, rather than using the self-dual and
anti-self-dual $SU(2)_{\pm}$ basis, we have used the $SO(4)$ basis,
and the appearance of $\epsilon_{abcd}$, instead of the
$\delta_{ac}\delta_{bd}$, takes care of the relative minus sign
between the $SU(2)_+$ and $SU(2)_-$. With the above given formulae,
the scalar potential can be directly obtained.


\end{document}